\documentclass[
 reprint,
 superscriptaddress,
 amsmath,amssymb,
 aps,
pra,
showkeys
]{revtex4-2}

\usepackage{soul}
\usepackage{float}
\usepackage{multibib}
\usepackage{hyperref}
\usepackage{xcolor}
\hypersetup{
    colorlinks,
    linkcolor={blue!50!blue},
    citecolor={blue!50!blue},
    urlcolor={blue!80!black}
}
\usepackage{siunitx}
\usepackage{import}
\usepackage{placeins}
\usepackage{xcolor}
\usepackage{lmodern}
\usepackage[braket, qm]{qcircuit}
\usepackage{graphicx}
\usepackage{dcolumn}
\usepackage{bm}
\usepackage{comment}
\usepackage[english]{babel}
\usepackage{blindtext}
\usepackage{physics}  

\DeclareMathOperator{\sinc}{sinc}

\begin{document}


\title{Efficient Light Propagation Algorithm using Quantum Computers}

\author{Chanaprom Cholsuk}
\email{chanaprom.cholsuk@uni-jena.de}
\affiliation{Abbe Center of Photonics, Institute of Applied Physics,
Friedrich Schiller University Jena, 07745 Jena, Germany}
\affiliation{Department of Computer Engineering, School of Computation, Information and Technology, Technical University of Munich, 80333 Munich, Germany}

\author{Siavash Davani}
\affiliation{Abbe Center of Photonics, Institute of Applied Physics,
Friedrich Schiller University Jena, 07745 Jena, Germany}
\affiliation{Max Planck School of Photonics, 07745 Jena, Germany}

\author{Lorcán O. Conlon}
\affiliation{Centre for Quantum Computation and Communication Technology, Department of Quantum Science, Australian National University, Canberra, ACT 2601, Australia}

\author{Tobias Vogl}
\affiliation{Abbe Center of Photonics, Institute of Applied Physics,
Friedrich Schiller University Jena, 07745 Jena, Germany}
\affiliation{Department of Computer Engineering, School of Computation, Information and Technology, Technical University of Munich, 80333 Munich, Germany}

\author{Falk Eilenberger}
\email{falk.eilenberger@uni-jena.de}
\affiliation{Abbe Center of Photonics, Institute of Applied Physics,
Friedrich Schiller University Jena, 07745 Jena, Germany}
\affiliation{Max Planck School of Photonics, 07745 Jena, Germany}
\affiliation{Fraunhofer Institute for Applied Optics and Precision Engineering IOF, 07745 Jena, Germany}

\date{\today}

\begin{abstract}
Quantum algorithms can potentially overcome the boundary of computationally hard problems. One of the cornerstones in modern optics is the beam propagation algorithm, facilitating the calculation of how waves with a particular dispersion relation propagate in time and space. This algorithm solves the wave propagation equation by Fourier transformation, multiplication with a transfer function, and subsequent back transformation. This transfer function is determined from the respective dispersion relation, which can often be expanded as a polynomial. In the case of paraxial wave propagation in free space or picosecond pulse propagation, this expansion can be truncated after the quadratic term. The classical solution to the wave propagation requires $\mathcal{O}(N log N)$ computation steps, where $N$ is the number of points into which the wave function is discretized. Here, we show that the propagation can be performed as a quantum algorithm with $\mathcal{O}((log{}N)^2)$  single-controlled phase gates,  indicating exponentially reduced computational complexity. We herein demonstrate this quantum beam propagation method (QBPM) and perform such propagation in both one- and two-dimensional systems for the double-slit experiment and Gaussian beam propagation. We highlight the importance of the selection of suitable observables to retain the quantum advantage in the face of the statistical nature of the quantum measurement process, which leads to sampling errors that do not exist in classical solutions.
\end{abstract}


\maketitle

\begin{figure*}[htp]
    \centering
    \includegraphics[width = 1\textwidth]{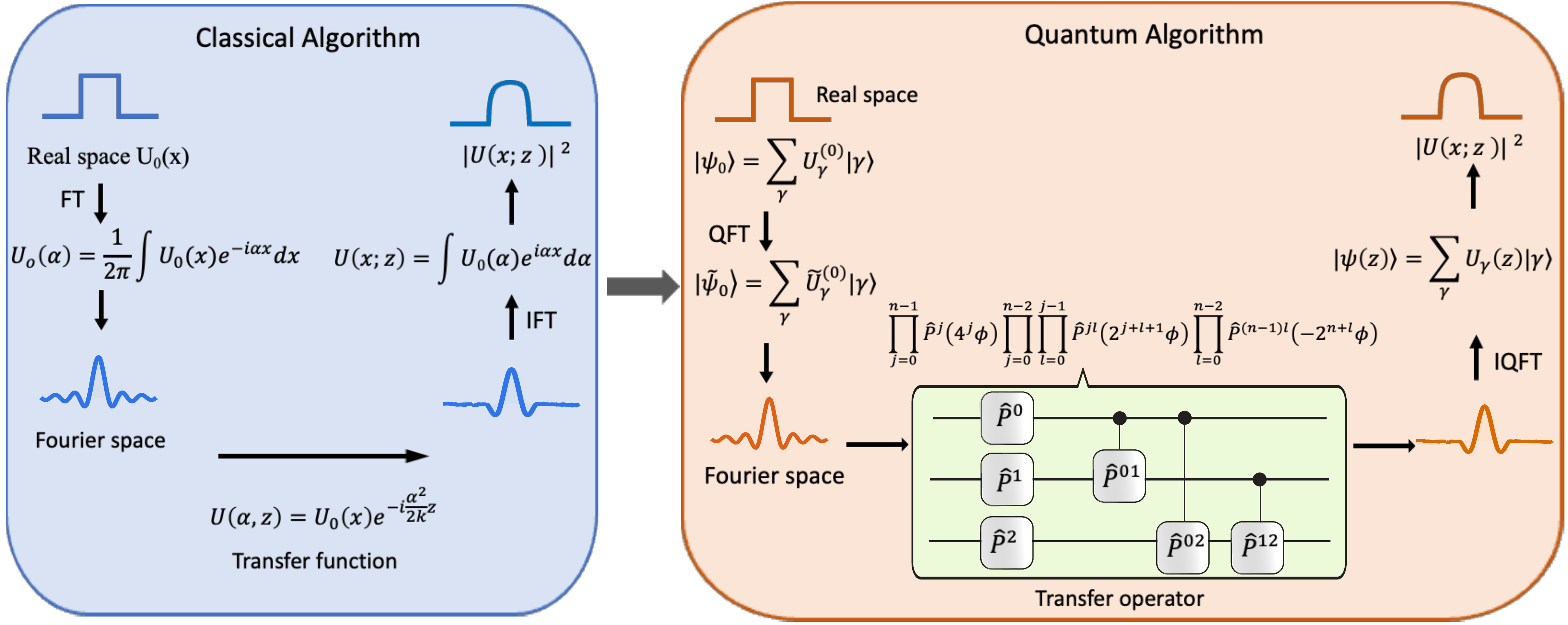}
    \caption{Schematic overview of classical and quantum propagation algorithm. In both cases, to solve the Helmholtz equation, we convert the initial wavefunction from real space to Fourier space via either the classical or quantum Fourier transform, then apply the transfer function, and finally take the inverse Fourier transformation to obtain the propagated wave in real space. The operator $P^j$ is indexed by the qubit $j$ and $P^{jl}$ is indexed by qubits $j$ and $l$ with the specific phase shift value.}
    \label{fig:overview}  
\end{figure*}

\section{Introduction}
Recently, major breakthroughs toward outstanding performance in computation, communication, and sensing by leveraging quantum properties have been demonstrated. Although many quantum systems are still under ongoing investigation \cite{Weber2010,Reilly2015,Pezzagna2021,Shaik2021,Cholsuk2022,Cholsuk2023,Cholsuk2024}, quantum computers \cite{Brien2007,Ladd2010} have potentially performed tasks or measurements, that cannot be done classically, e.g.,\ in quantum sensing \cite{Conlon2023,Glenn2018}, quantum communication \cite{Clivati2022,Lorcan2023}, and interferometry \cite{Vogl2021}.\\
\indent Quantum algorithms in particular have been expedited to overcome the boundary of the problems unsolvable by classical computation \cite{Centrone2021, Zhang2021, Conlon2023} and also speed up classical algorithms \cite{Arute2019, Zhong2020, Madsen2022}. Recent accomplishments in wave physics have e.g. demonstrated quantum advantage for solving the Poisson equation \cite{Cao2013}, computing the electromagnetic scattering cross sections \cite{Clader2013}, and simulating non-classical interference phenomena \cite{Costa2019}. Some of these have demonstrated exponential speedup over the best-known classical algorithms, highlighting the efficiency with which quantum algorithms can treat interference problems.\\
\indent Among the most prominent quantum algorithms are those which leverage the quantum Fourier transform (QFT) \cite{Coppersmith1994AnAF,doi:10.1098/rspa.1998.0163} to efficiently convert between real space and Fourier space representations. These include quantum phase estimation (QPE) \cite{doi:10.1098/rspa.1998.0164}, and Shor's algorithm \cite{RevModPhys.68.733} among many others. From a physical point of view, Fourier transformations are a particularly powerful tool to solve linear differential equations, such as those encountered in wave propagation. Hence, solving wave propagation problems in quantum computers is an active area of research, as can be seen from recent works, consisting of wave equations under Dirichlet and Neumann boundary conditions \cite{Costa2019} and the single-particle Schr\"{o}dinger equation with various potential barriers \cite{Abouelela2020, Bouten2019}.\\ 
\indent Nevertheless, solving the wave propagation for modern optics remains unrealized so far. Unlike prior studies of wave propagation by quantum computers, we herein focus on solving wave propagation from the Helmholtz equation by implementing a Fourier transformation technique inspired by the classical beam propagation method (BPM). Classically, the BPM solves the Helmholtz equation by intermediate transfer of the wavefunction, which is discretized into $N$ points in each dimension, into the Fourier domain using the Fast Fourier Transformation (FFT), the multiplication by a specific transfer function, and subsequent back transformation into real space. By implementing this procedure in quantum computers, we can exploit QFT and an efficient quadratic phase operator, introduced in this work, to solve the Helmholtz equation in paraxial approximation, achieving computational complexity speed up compared to the classical BPM.\\
\indent In addition, in this work, the difference between classical and quantum algorithms is demonstrated and compared. It is shown that only $n=\log_{2}N$ qubits are required to store the complete wavefunction, and we propose a method to implement the transfer functions whose the dispersion relation has been expanded up to the order of $p$-th with $\mathcal{O}((\log{}N)^p)$ operations efficiently using a gate-based quantum computation algorithm. This implies the generalization of our QBPM being able to apply for an arbitrary light propagation. For demonstration purposes, we used Qiskit \cite{Qiskit} to implement the QBPM and simulated it in a noiseless environment, i.e. on a quantum computer simulator. We showcase paraxial wave propagation ($p = 2$), in both one and two-dimensional environments. That is, we simulated the double slit experiment for one-dimensional propagation and Gaussian beam propagation in two lateral dimensions.

\section{Classical wave propagation} \label{sec:general_prop}
The propagation of electromagnetic, monochromatic waves  in free space or in an isotropic media is described by the Helmholtz equation,
\begin{equation}
    \nabla^2 U(\textbf{r},\omega)+k^2(\omega)U(\textbf{r},\omega) = 0, \label{eq:propagation}
\end{equation}
where $U(\textbf{r},\omega)$ is an arbitrary field and $\textbf{k}(\omega)$ is a wave vector, which is based on some dispersion relation $\omega=\omega(\textbf{k})$. A typical problem in wave propagation is finding the solution of the wave field  $U(\textbf{r}_\perp{}; z)$ given a specific wave frequency $\omega$ and a known initial value $U_0(\textbf{r}_\perp)$ = $U_0(\textbf{r}_\perp;z=0)$ defined along the transverse coordinates $\mathbf{r}_\perp$, which are, $\mathbf{r}_\perp=x\mathbf{e}_x$ in case of one transverse dimension and  $\mathbf{r}_\perp=x\mathbf{e}_x+y\mathbf{e}_y$  in case of two.\\
\indent One common approach for solving this problem is to first take the Fourier transform of the initial field with respect to the transverse coordinates \cite{photonics} as depicted in Fig.\,\ref{fig:overview}. In case of a problem with one transverse dimension, this is given as: 
\begin{equation}
  \tilde{U}_0(\alpha)=\frac{1}{2\pi}\int_{-\infty}^{\infty}U_0(x)e^{-i\alpha x}dx, 
\end{equation}
where $\alpha$ is the transverse wave number in $x$ direction.  To attain the propagated field in the frequency domain, we need to take the transfer function into account. If the transverse wave numbers are much smaller than the longitudinal wave numbers, i.e. $\alpha^2 \ll k^2$, the transfer function can be expressed by the quadratic polynomial approximation of the dispersion relation, i.e.
\begin{eqnarray}
    H(\alpha;z,k) &=&  e^{i\sqrt{k^2 - \alpha^2}z} \nonumber \\
     &\approx& e^{ikz}e^{-i\frac{\alpha^2}{2k}z}.
\end{eqnarray}
The propagated field in the frequency domain is then given by
\begin{equation}
    \tilde{U}(\alpha;z)\approx \tilde{U}_0(\alpha)e^{ikz}e^{-i\frac{\alpha^2}{2k}z},
    \label{eq:propagated_field}
\end{equation}
where the first exponent is just a fixed phase and can be ignored for many problems. Likewise, if we consider the dispersion of a temporal pulse, we can expand the dispersion relation into a Taylor series and truncate after the quadratic term, if the pulse is not too short (typically sub-picosecond) and the linear term can be removed by the introduction of a co-moving frame of reference. Finally, we take the inverse Fourier transform to obtain the propagated field in space $U(x;z)$. The generalization of this to two or more transverse dimensions is straightforward.\\
\indent For a numerical solution, we discretize the wavefunction $U(x)$ on $N$ equidistant points in space $x_\gamma=\gamma\Delta{}x$ with $\gamma\in [-N/2,...,0,1,...,N/2-1]$, such that $U^{(0)}_\gamma=U_0(\gamma\Delta{}x)$ where $U^{(0)}_\gamma$ is the discretized wavefunction. $N=2^n$ is usually selected as a power of two, consistent with the number of qubits $n$ in the quantum algorithm. The Fourier transformation with the FFT algorithm requires $\mathcal{O}(N \log(N))$ steps and yields the discretized Fourier amplitudes $\tilde{U}^{(0)}_\gamma$ at the frequencies $\alpha_\gamma=\gamma\Delta\alpha$, where $\Delta\alpha=2\pi/(N\Delta x)$.\\
\indent The solution then requires the multiplication of the transfer function for each discrete frequency $H_\gamma=H(\alpha_\gamma;z,k)$, which requires $\mathcal{O}(N)$ steps, yielding the Fourier amplitudes $\tilde{U}_\gamma(z)$  and a subsequent application of the inverse FFT to calculate $U_\gamma(z)$. The overall cost of applying the algorithm is thus $\mathcal{O}(N \log(N))$.\\
\indent It should be noted that the algorithm does not find a perfect solution to the Helmholtz equation, because its discretization in real and Fourier spaces imply periodic boundary conditions in both spaces, yielding a non-vanishing discretization error. If an analytic solution $I_\mathrm{Ref}(x_\gamma;z)$, or any other reference solution for a particular initial field is known, we can measure the impact of the error on a wavefunction's intensity using e.g.\,a root-mean-square-error $\epsilon(z)$:
\begin{equation}
 \epsilon(z) = 
\sqrt{\frac{\sum_\gamma\left(I_\mathrm{Ref}(x_\gamma;z)-I_\gamma(x_{\gamma};z)\right)^2}
{\sum_\gamma I_\mathrm{Ref}(x_\gamma;z)}},
\end{equation}
where $I_\mathrm{Ref}(x_\gamma;z)=|U_\mathrm{Ref}(x_\gamma;z)|^2$ is the absolute value square of the analytic solution and $I_\gamma(z)=|U_\gamma(x_{\gamma};z)|^2$ is the absolute value square of the numerically solved discrete solution.\\
\indent We shall later see that the QBPM introduces random errors in the measurement process and hence $\epsilon(z)$ is a statistically distributed quantity in its own right. Its distribution can be sampled over many repetitions $N_\mathrm{sim}$ and described by its mean error $\mu=\expval{\epsilon}$ and the standard error $\sigma=(\expval{\epsilon^2}-\expval{\epsilon}^2)^{1/2}$.
\begin{figure*}[htp]
    \centering
    \includegraphics[width = 1\textwidth]{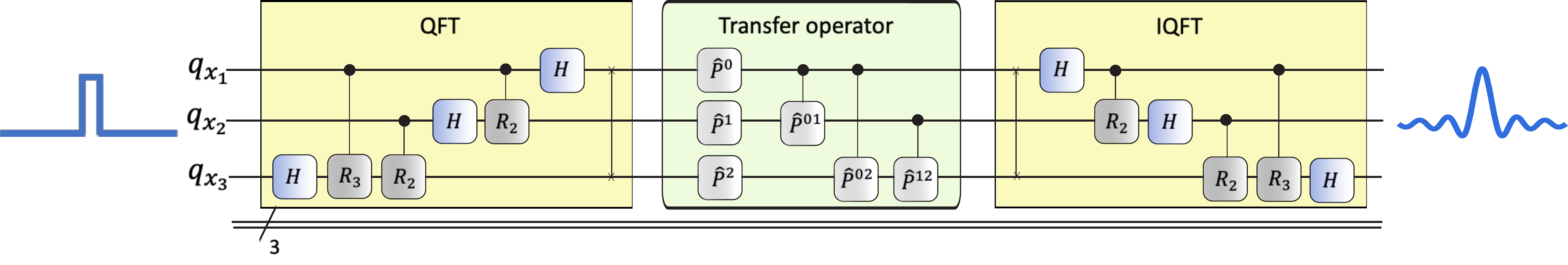}
    \caption{One-dimensional QBPM logic circuit for the double-slit experiment.}
    \label{fig:1d-circuit}
\end{figure*}
\section{Quantum algorithm for propagation} \label{sec:algorithm}
To formulate the quantum algorithm for the wave propagation discussed above, we map the classical BPM steps onto quantum gates. All procedures are illustrated in Fig.\,\ref{fig:overview}. The step-by-step derivation for our proposed algorithm is as follows.
\begin{enumerate}
    \item Selection of qubits: we create a wavefunction $\ket\psi$ from $n=\log N$ qubits, the state of which can be described by an arbitrary superposition of its computational basis states $\ket{\gamma}$ with $\gamma\in [-N/2,...,0,1,...,N/2-1]$, which can be represented as a binary number $\gamma= -a_{n-1}2^{n-1} + \sum_{j=0}^{n-2}a_j2^j $ composed of its binary digits $a_j\in{\{0,1\}}$ using the two's complement to indicate negative values. Mapping the $N$ point wavefunction onto the $n$ qubits already entails exponential enhancement of memory efficiency.
    \item Initialization: We represent the initial wavefunction $U^{(0)}_\gamma$ as $\ket{\psi_0}=\sum_\gamma{U^{(0)}_\gamma\ket{\gamma}}$. An efficient and general solution to this problem is a subject of active research and not discussed in this work. However, for various special cases, some specific solutions have been shown \cite{Schuld2021}. For the frequently encountered case of Gaussian beams, we can, e.g., make use of the Kitaev-Webb algorithm \cite{Kitaev2008}. 
    \item Fourier domain: We perform a standard QFT on $\ket{\psi_0}$ transforming the state into $\ket{\tilde{\psi_0}}=\sum_{\gamma}{\tilde{U}^{(0)}_\gamma\ket{\gamma}}$. Note that $\gamma$ now indicates the wave number index. As shown in Fig.~\ref{fig:1d-circuit}, $H$ is the Hadamard gate, followed by the phase gate $R_{j}$, which leads to phase shift by the following expression; $R_{j}(\phi)=\begin{bmatrix}
1 & 0 \\
0 & \mathrm{exp}(2\pi i/2^{j}) 
\end{bmatrix}$.
    \item Second-order propagation: In the Fourier space, we need to multiply each $\tilde{U}^{(0)}_\gamma$ term with the transfer function $e^{-i\frac{\alpha_\gamma^2}{2k}z}=e^{i\phi\gamma^2}$ with $\phi=-\frac{2\pi^2 z}{N^2\Delta x^2k}$. This represents a phase shift, which only depends on the fixed value of $\phi$ and the square of the state index $\gamma$. An efficient implementation can be done by rewriting $\gamma$ in its binary form. This leads to (See details in Supplementary Material S1):
    \begin{eqnarray}\label{eq:gamma_2_decompos}
        \gamma^2 &=& \sum_{j=0}^{n-1}{4^ja_j}
        +\sum_{j=0}^{n-2}\sum_{l=0}^{j-1}{2^{j+l+1}a_ja_l} - \sum_{l=0}^{n-2}{2^{n+l}a_{n-1}a_l}. \nonumber \\
    \end{eqnarray}
    Note that $j$ and $l$ are qubit indices and that either $a_j=1$ or $a_j=0$ for all $j$.\\
\indent This allows for an efficient implementation of the transfer function operator, as we note the summands of the first sum have the value $4^j$ under the condition that qubit $j$ is in the $\ket{1}$ state and zero otherwise. Likewise, all summands of the middle sum have the value $2^{j+l+1}$ under the condition that the qubit-pair $j$ and $l$ are both in their $\ket{1}$ state and zero otherwise. The last term is for two's complement and of identical structure.\\
\indent The multiplication of the transfer operator $\hat{U}$ can thus be decomposed into a series of phase gates and single-controlled phase gates, such that $\ket{\tilde{\psi}(z)}=\hat{U}\ket{\tilde{\psi_0}}$.
\begin{equation}
\hat{U}=\prod_{j=0}^{n-1}\hat{P}^j(4^j\phi)\prod_{j=0}^{n-2}\prod_{l=0}^{j-1}\hat{P}^{lj}(2^{j+l+1}\phi)\prod_{l=0}^{n-2}\hat{P}^{l(n-1)}(-2^{n+l}\phi). \label{eq:transfer}
\end{equation}
\indent Here, the phase gates are defined as $\hat{P}^j(\phi)=\begin{bmatrix}
1 & 0 \\
0 & \mathrm{exp}(i\phi) 
\end{bmatrix}$, where the $j$ index indicates that it acts on the $j$-th qubit. Likewise, the controlled phase gates are defined using $\hat{P}^{lj}(\phi)=\begin{bmatrix}
1 & 0 \\
0 & \mathrm{exp}(i\phi) 
\end{bmatrix}$, where the control and target qubits are respectively the $l$-th and $j$-th qubits. Note that the phases that each phase gate or controlled phase gate apply correspond to the coefficients in the two's-complement decomposition in Eq.\,\ref{eq:gamma_2_decompos}. The matrix representation of these gates can be seen in Supplementary Material S2.
\item The calculation of the final state $\ket{\psi(z)}$ is done by subsequent application of the inverse QFT (IQFT).
\end{enumerate}

\indent By measuring the final state, we collapse the field into a computational basis state with a probability of collapse being equal to the absolute value square of the wavefunction at this point in space, hence its intensity:
   \begin{equation}
        p(x=x_\gamma)=|\bra{\gamma}\ket{\psi(z)}|^2 \propto I_\gamma(z).
    \end{equation}
\indent This is equivalent to measuring a single photon in the final wavefield on a camera with $N$ pixels. The complete intensity distribution can be sampled from a series of $N_s$ ``single photon clicks", termed ``shots". If the intensity distribution is not required, but some other measurable is sought-after, then a different approach for measurement may be better suited.\\
\indent Note, however, that any kind of sampling induces a sampling error, with - as opposed to the discretization error - is statistical in nature. The impact of both types of error can hence be separated from an ensemble average by calculating the statistical features of the RMSE with a fixed number of shots $N_s$ over many simulations $N_\mathrm{sim}$. If the sample based estimator is unbiased then a mean error $\mu$ is indicative of the discretization error and the standard error $\sigma$ is indicative of sampling noise.\\
\indent Both QFT as well as the transfer function multiplication operation require $\mathcal{O}(n^2)$ operations, thus the entire calculations scales with  $\mathcal{O}((\log N)^2)$. As such, the QBPM is exponentially faster than the classical BPM algorithm.\\
\indent Although all above BPM and QBPM are derived based on the second-order approximation, it is crucial to note that our algorithm can also be applied for higher order polynomials $\alpha^p$, as are required to e.g. model pulse propagation in dispersive media or to go beyond the paraxial approximation for large-angle wave propagation. A general formula can be found in the Supplementary Material S1 and it can be seen that the speedup is still maintained, as the scaling for these operations is $\mathcal{O}(n^p)$.

\section{One-dimensional propagation} \label{sec:1d}
\indent We now turn to implementing the above algorithm and presenting the double-slit experiment as a test case to validate the method's functionality and to investigate the relative impact of discretization error $\mu$, which is a generic feature of the discretized solver, vs.\, the sampling noise $\sigma$, which is a feature of the measurement procedure inherent in quantum computers.\\
\indent The initial state was chosen as:
\begin{equation}
    U_{0}(x) = \frac{1}{\sqrt{2w}}
                    \begin{cases} 
                        1 & |x-\frac{d}{2}|\leq \frac{w}{2}, \quad |x+\frac{d}{2}|\leq \frac{w}{2} \\
                        0 & \text{Otherwise}.
                    \end{cases}
\end{equation}
Where the separation of the slits $d$ is 0.5 mm, the width of each slit $w = 0.1$ mm, and the wavelength of the monochromatic light $\lambda = 532 $ nm. The number of qubits is 15 qubits.\\
\indent The intensity pattern at the output is a function of $z$ and the analytical solution is \cite{book_optics}:
\begin{equation}
    I_\mathrm{ref}(x;z) = I_0
    \cos^2\left(\pi\frac{d}{\lambda}\sin\theta\right)
    \sinc^2\left(\pi\frac{w}{\lambda}\sin\theta\right),
\end{equation}
where $\tan\theta=x/z$ and $I_0$ is the maximum intensity in the center.
\begin{figure}[htp]
    \centering
    \includegraphics[width = 0.5\textwidth]{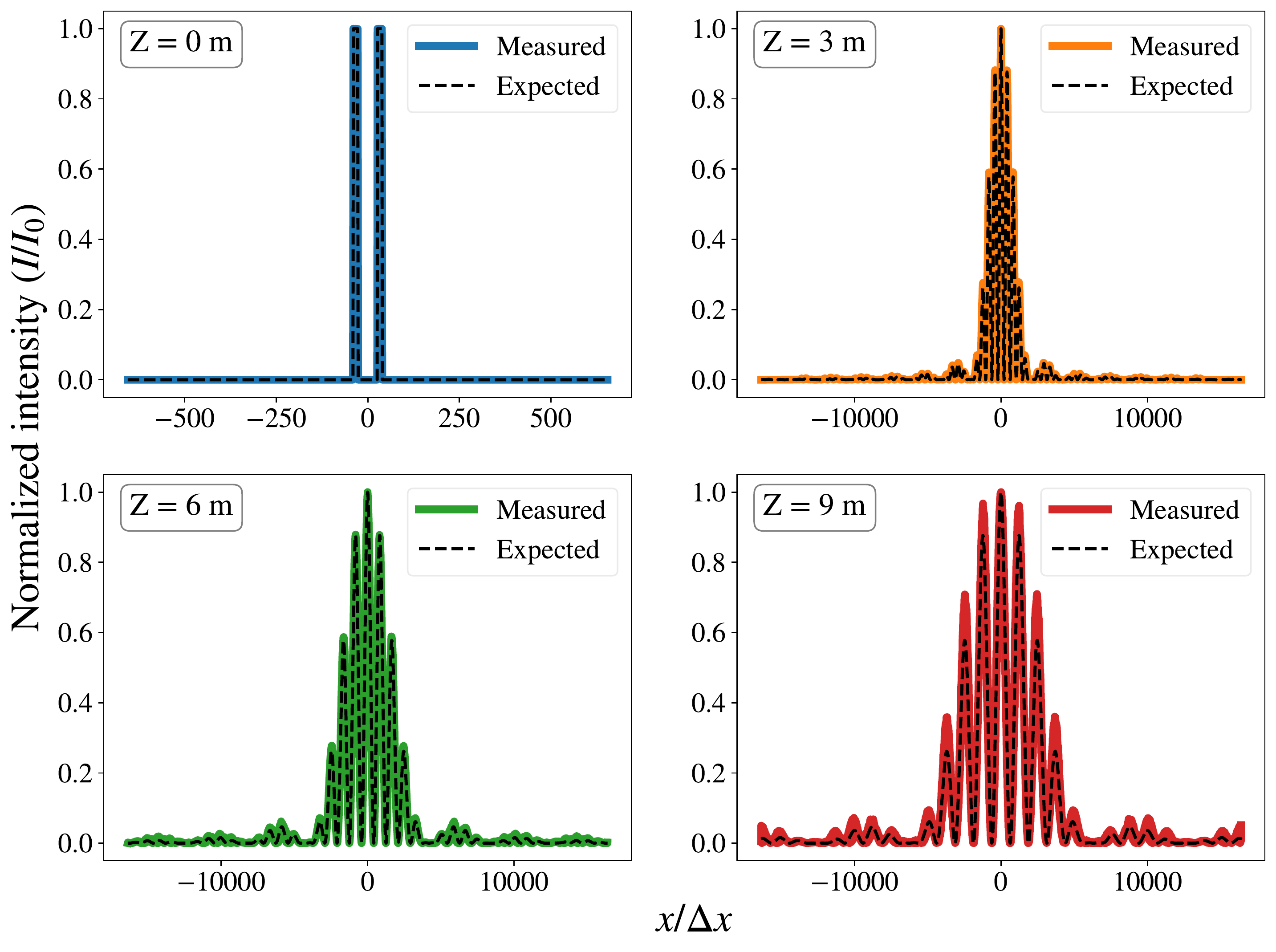}
    \caption{Interference pattern sampled from $N_s=100,000$ shots and 15 qubits as a result of simulating the double-slit experiment using the QBPM in the 1D case compared to the analytical solution with different propagation distances.}
    \label{fig:slit}
\end{figure}
\begin{figure*}
    \centering
    \includegraphics[width = 0.7\textwidth]{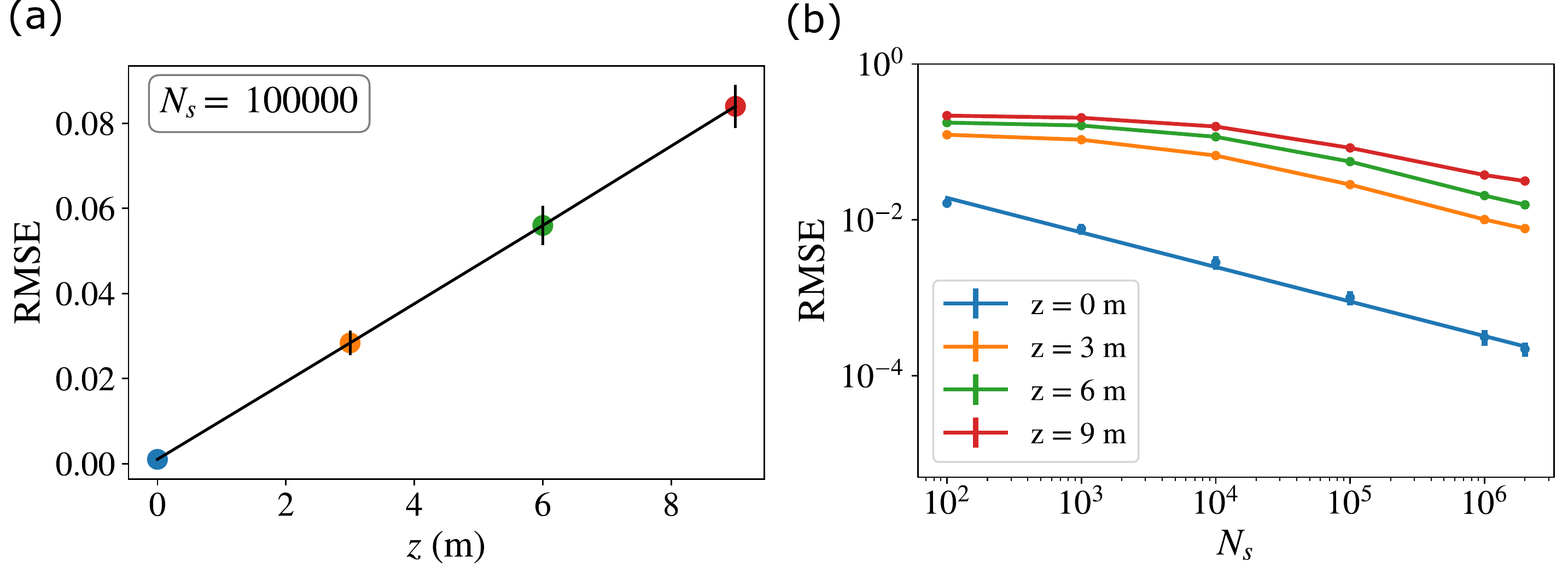}
    \caption{Mean error $\mu$  (points) and standard error  $\sigma$  (bars) are accumulated from $N_\mathrm{sim}=100$ simulations. (a) As a function of propagation distance $z$ where the number of shots $N_s$=100,000. (b) As a function of $N_s$ at different propagation distances.}
    \label{fig:RMSE-1D}
\end{figure*}
\subsection{Simulated interference pattern and Error analysis}
\indent We now compare our quantum algorithm to the classical analytic solution for the double-slit experiment. As depicted in Fig.\,\ref{fig:slit}, we can see the result of the quantum simulation after different $z$ propagation distances compared to the analytical solution. The result agrees well with the expected interference pattern.\\
\indent Both discretization error $\mu$ and sampling error $\sigma$ depend on the propagation distance $z$ and the number of shots $N_s$. In Fig.\,\ref{fig:RMSE-1D}(a), both are plotted as a function of the propagation distance. The result indicates that the discretization error increases linearly with the propagation distance, and that it is larger than the sampling error except for $z=0$. This is an important finding, as it naturally gives a rise to a selection rule for the number of shots $N_{s,max}$, beyond which a reduction of the shot noise will not improve overall precision, beyond the systematic discretization error.\\
\indent For the given simulation, such a number can be estimated from the data shown in Fig.\,\ref{fig:RMSE-1D}(b). While at $z=0$ the total error can be reduced with the inverse of the square root of $N_s$ to an arbitrarily low value, a minimal error is evident for all $z>0$.  For all investigated distances this likely occurs in our simulation at $N_{s,max}\approx10^6$. As a complete reconstruction of the absolute value of the wavefunction by means of sampling does not appear to be a suitable mode of operation for a quantum simulator anyway, a more systematic investigation of this boundary is not undertaken. \\
\begin{figure*}
    \centering
    \includegraphics[width = 1\textwidth]{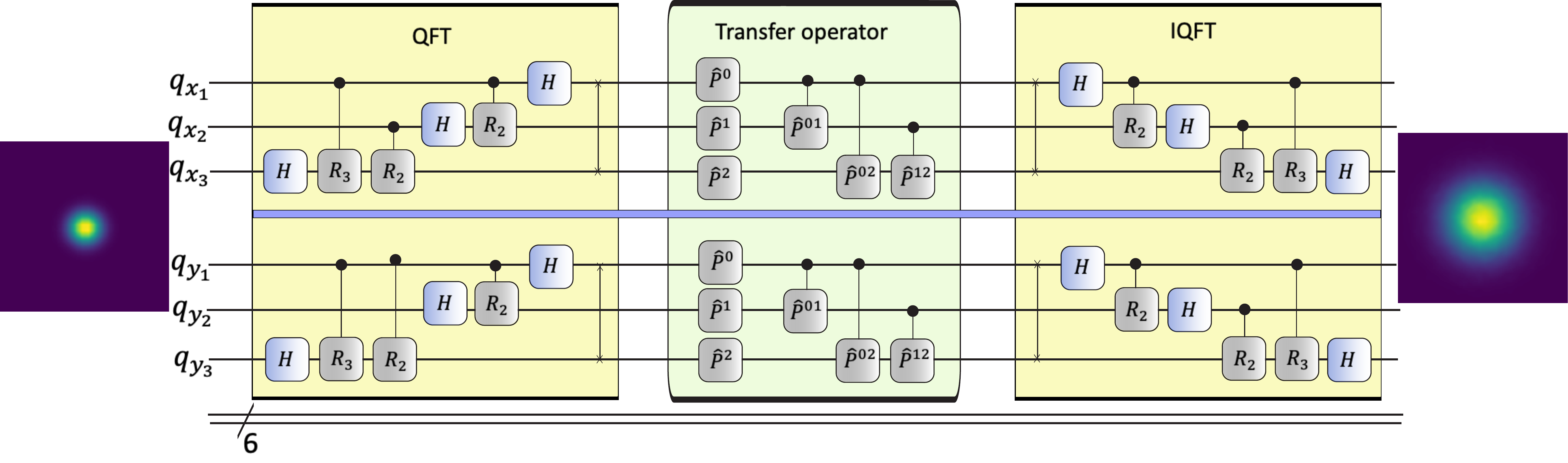}
    \caption{Two-dimensional QBPM logic circuit for Gaussian beam propagation in free space where the upper part of the circuit is for propagation in $x$ direction and the lower part is for $y$ direction.}
    \label{fig:circuit_2d}
\end{figure*}
\section{Two-dimensional propagation} \label{sec:2d}
\indent In addition to the simulation of one-dimensional propagation as demonstrated in Sec.\,\ref{sec:1d}, we herein extend our propagation algorithm to two-dimensional propagation and select a Gaussian beam propagation in free space as a case study.
\subsection{Computational model}
\begin{figure}
    \centering
    \includegraphics[width = 0.49\textwidth]{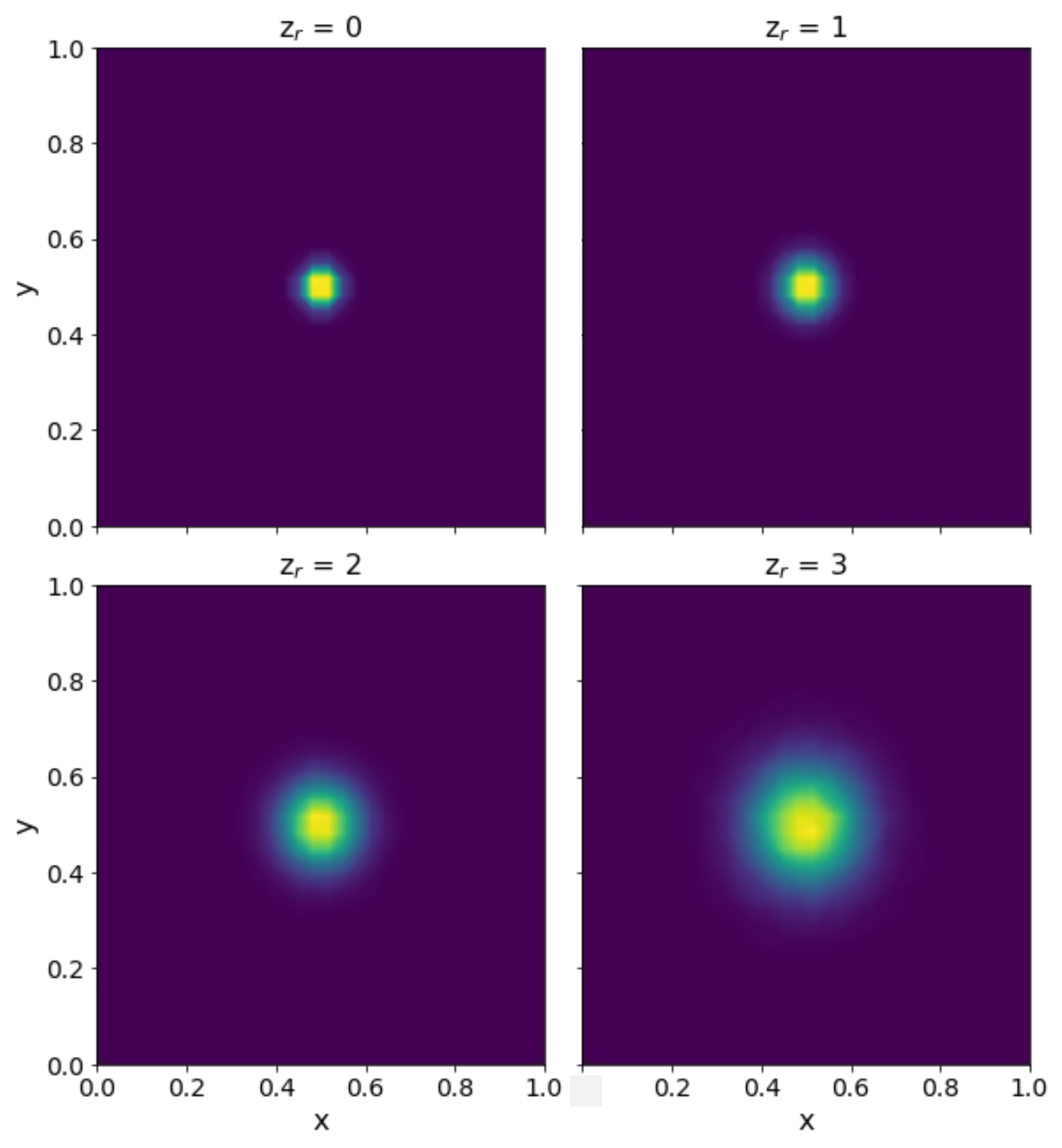}
    \caption{A Gaussian beam propagated by distance's ratio (a) $z_{r} = 0.0$, (b) $z_{r} = 1.0$, (c) $z_{r} = 2.0$, and (d) $z_{r} = 3.0$, where the propagation distance ($z$) = $z_{r}z_0$ and the Rayleigh length $z_0$ = $\frac{kw_0^2}{2}$. The number of shots (N$_s$) is 50,000 shots with a single run. The number of qubits is 5 qubits for each $x$ and $y$ coordinates.}
    \label{fig:prop2D}
\end{figure}

 The wave function is initialized  by
\begin{equation}
    U^{(0)}_{\gamma,\zeta} = A_0e^{-\frac{(x_\gamma-x_0)^2+(y_\gamma-y_0)^2}{w_0^2}}, \label{eq:initial2d}
\end{equation}
where $A_0$ is the normalization constant and $w_0$ is the beam waist. As such, here, $\alpha$ and $\beta$ are discretized by $2\pi\gamma/\Delta xN$ and $2\pi\zeta/\Delta yN$, accordingly. In this example, we define the width at focal plane ($w_0$) at 0.05 m, wavelength ($\lambda$) at 532 nm, the number of qubits in each direction set to 5 qubits (10 qubits total), and the Gaussian is centered at the domain's center, such that $x_0=y_0=0$.\\
\indent For two-dimensional propagation, we can straightforwardly add one more coordinate into the circuit as the algorithm is separable along $x$ and $y$ directions. This also applies to non-separable input states, which, however, have not been demonstrated in this work. This implies that all steps remain the same as those of the one-dimensional case, hence the quantum algorithm will be duplicated for the other axis as depicted in Fig.\,\ref{fig:circuit_2d}. It should be noted that the dimensional independence is a feature of the dimensional independence of diffraction from the quadratic polynomial approximation and not of the quantum algorithm.\\ 
\indent The Gaussian beam is propagated with the propagating distance ($z$) as illustrated in Fig.\,\ref{fig:prop2D}. It is obvious that when $z_r$ is larger, the Gaussian beam broadens, as expected.\\ 
\subsection{Observable and Error analysis}
\indent As opposed to the 1D-case, we here aim to highlight the impact of a well-selected observable on the number of shots $N_s$ required for a good estimation of some observable. As a test case, we estimate the beam waist $w(z)$ as shown in Eq.\,\ref{eq:waist_qc} as a function of the propagation distance,
\begin{equation}
    w_Q=\sqrt{\sum_{i=0}^{N-1}\sum_{j=0}^{N-1}\left((x_i-x_0)^2 + (y_j-y_0)^2 \right)P_{ij}},
    \label{eq:waist_qc}
\end{equation}
where $x_i$ and $y_j$ run from 0 to the position domain ranged $N$. Where at each position ($x_{i},y_j$) we have measured the number of times the circuit collapses into this final state $P_{ij}=N_{ij}/N_\mathrm{s}$. We compare it with the value obtained from classical discretized solution as shown below,
\begin{equation}
    w_\mathrm{ref}=\sqrt{\sum_{i=0}^{N-1}\sum_{j=0}^{N-1}\left((x_i-x_0)^2 + (y_j-y_0)^2 \right)|U_{ij}|^{2}},
\end{equation}
where $|U_{ij}|$ is the absolute value of the wavefunction. As opposed to the last section we here reference against the discrete approximation; hence the difference $\epsilon=w_Q-w_\mathrm{ref}$ is exclusively caused by the sampling error. Again, we measure the standard error $\sigma_w=(\expval{\epsilon^2}-\expval{\epsilon}^2)^{1/2}$. Both values depend on the number of shots $N_s$ and the propagation distance $z$ and have been determined from $N_\mathrm{sim}$=100  independent simulations.\\
\begin{figure}[htp]
    \centering
    \includegraphics[width = 0.35\textwidth]{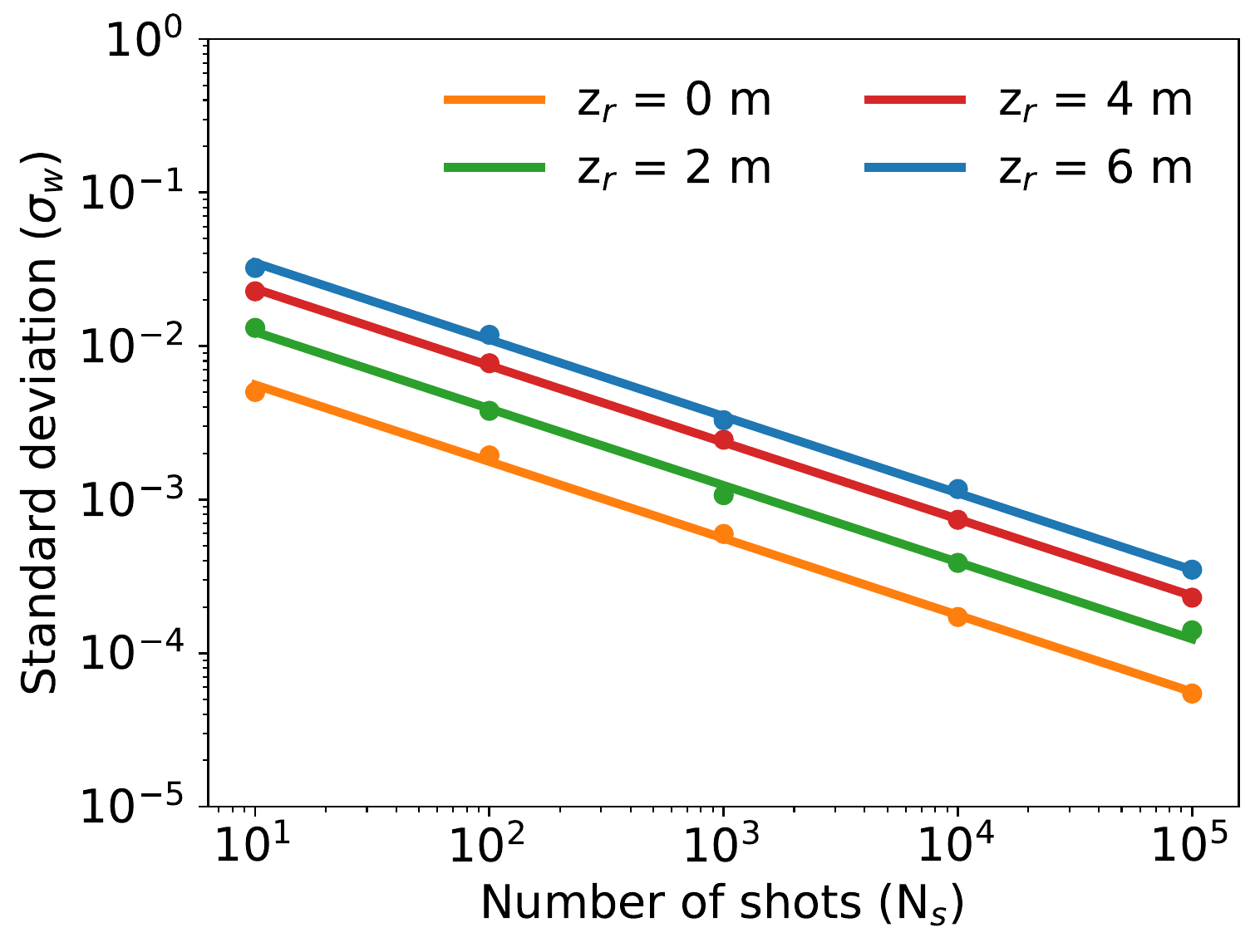}
    \caption{Sampling error (measured by the standard error $\sigma_w$) of the measurable $w_Q$, i.e. the radius of the gaussian wave packet, determined over $N_\mathrm{sim}=100$ as  a function of the propagation distance $z_r$ and a function of the number of shots $N_s$. Even for longer propagation lengths, $w_Q$ is a very good approximation of $w_c$}
    \label{fig:error-2D}
\end{figure}

\indent We now analyze the size of the standard error as a function of the propagation length and the number of shots. This data is displayed in  Fig.\,\ref{fig:error-2D}. The data indicates that only $N_s=100$ already allows for a very good sampling precision of the beam waist radius. If a better approximation is required then the number of shots can be increased, and we observe an increase in precision of~1/$\sqrt{N_s}$. Larger propagation distances inherit the more considerable error, consistent with the 1D case. This shows that the fact, that quantum computers require sampling, does not systematically ill-affect the efficiency of the algorithm; if a sufficiently well-behaved measureable is required, which we argue is a common use case, a very moderate number of shots is sufficient for a good estimation of the measurable.\\
\indent Nonetheless, as the propagation in the 2D case is more sophisticated than that in 1D, we also benchmarked our QBPM among four different cases, consisting of noisy quantum simulator, noiseless quantum simulator, classical discretized algorithm, and classical analytic solution to guarantee the accuracy of our algorithm. All of these comparisons agree well as can be seen in Supplementary Material S3.

\section{Conclusion} \label{sec:conclusion}
We have developed a quantum beam propagation algorithm based on the paraxial approximation. We show that our quantum algorithm can efficiently simulate beam propagation in both one and two dimensions by implementing only phase gates and controlled phase gates with an exponential speedup over the classical solution, based on the fast Fourier transformation. This allows for a low number of qubits and operations used for the propagation when compared with the classical counterpart. We also carry out a rigorous error analysis to give more insights into the role of the sampling error, which is a new source of errors, when quantum computers are used. We analyze two scenarios: the sampling of the absolute value squared of the wavefunction and find that the sampling error falls below the discretization error after only a few thousand shots $N_s$. In a second scenario, we only analyze a scalar measurable, in this case, the waist of a broadening gaussian beam. Here we find that a very good accuracy can be achieved with only a few hundred shots and hence the quantum advantage can be maintained for a broad range of scenarios.\\
\indent As the 2D Helmholtz equation can have factorized and non-factorized initial states, the demonstration for this initialization problem can fulfill this important gap. In principle, this will allow one to simulate a system bounded by Dirichlet or Neumann boundary conditions where its initial states are not separable.\\
\indent We further argue that our work does not merely add one more algorithm to the toolbox of quantum simulation. Instead, it bridges the gap between Fourier optics and quantum computation and lays the foundation for quantum computers as a tool to efficiently simulate more complex problems in photonics, such as Boson sampling or spontaneous conversion in photonic nanostructures.

\begin{acknowledgments}
This work was funded by the Federal Ministry of Education and Research (BMBF) under grant numbers 13XP5053A (NanoscopeFutur-2D) and 13N16292 (ATOMIQS). The authors received funding from the Deutsche Forschungsgemeinschaft (DFG, German Research Foundation) - CRC NOA 1375, B3 and Projektnummer 445275953. This research was supported by the German Space Agency DLR with funds provided by the Federal Ministry for Economic Affairs and Climate Action BMWK under grant numbers 50WM2165 (QUICK3).\,C.C. acknowledges a Development and Promotion of Science and Technology Talents Project (DPST) scholarship by the Royal Thai Government. We acknowledge the use of IBM Quantum services for this work. The views expressed are those of the authors, and do not reflect the official policy or position of IBM or the IBM Quantum team.
\end{acknowledgments}


\bibliography{main-arxiv}



\end{document}